\documentstyle[12pt,epsfig]{article}
\makeatletter
\def\vereq#1#2{\lower3pt\vbox{\baselineskip1.5pt \lineskip1.5pt
\ialign{$\m@th#1\hfill##\hfil$\crcr#2\crcr\sim\crcr}}}
\makeatother
\begin{document}

\begin{titlepage}
\begin{flushright}
\end{flushright}

\vskip 1cm
\begin{center}
{\large\bf A simple model of generating fermion mass hierarchy \\
in $N=1$ supersymmetric $6D$ $SO(10)$ GUT}
\vskip 1cm {\normalsize Naoyuki Haba$^{1,2}$, and Yasuhiro Shimizu$^2$}
\\
\vskip 0.5cm {\it $^1$Faculty of Engineering, Mie University, Tsu, Mie,
  514-8507, Japan}\\
\vskip 0.5cm {\it $^2$Department of Physics, Nagoya University, Nagoya, 
  464-8602, Japan}
\end{center}
\vskip .5cm
\begin{abstract}

We suggest simple models which produce the suitable fermion
 mass hierarchies and flavor mixing angles based on the 
 6 dimensional $N=1$ supersymmetric $SO(10)$ grand unified theory 
 compactified on a $T^2/(Z_2 \times Z_2')$ orbifold.
We introduce 6D and 5D matter fields, which contains
 the 1st and 2nd generation matter fields as the zero modes,
 respectively. 
The 3rd generation matter fields are
 located on a 4D brane. 
The Yukawa couplings for bulk fields
 are suppressed by volume factors from extra dimensions.
The suitable fermion mass hierarchies and flavor mixings 
 are generated by the volume suppression factors.

\end{abstract}
\end{titlepage}
\setcounter{footnote}{0}
\setcounter{page}{1}
\setcounter{section}{0}
\setcounter{subsection}{0}
\setcounter{subsubsection}{0}

\section{Introduction}
\label{sec:intro}

Grand unified theories (GUTs) are very attractive models in which the three
 gauge groups are unified at a high energy scale. 
However, one of the most serious problems to construct a model of GUTs 
 is how to realize the mass splitting  between the triplet and 
 the doublet Higgs particles in the Higgs sector. 
This problem is so-called triplet-doublet (TD) splitting problem. 
A new idea for solving the TD splitting problem 
 has been suggested in higher dimensional 
 GUTs where the extra dimensional 
 coordinates are compactified on 
 orbifolds\cite{5d}-\cite{so(10)others}.
In these scenarios, 
 Higgs and gauge fields are propagating in
 extra dimensions, and  
 the orbifolding realizes the gauge group
 reduction and the TD splitting since  
 the doublet (triplet) Higgs fields have (not)
 Kaluza-Klein zero-modes. 
A lot of attempts and progresses have been done in 
 the extra dimensional GUTs on orbifolds\cite{6d}-\cite{Shafi:2002ii}.
Especially, the reduction of $SO(10)$ gauge symmetry 
 and the TD splitting solution are 
 first considered in 6D models in Refs.~\cite{ABC,HNOS}.

As for producing fermion mass hierarchies,  
 several trials have been done 
 in the extra dimensional GUTs on 
 orbifolds\cite{HSSU,HKS,HS,flavor,Hebecker:2002re,comple,
SO(10)SU(6)gauge,Hebecker:2002xw,Shafi:2002ii}.
The model in Ref.~\cite{HKS,HKS2,HS} can induce the natural
 fermion mass hierarchies and flavor mixings 
 by introducing extra vector-like 
 generations 
 which propagate 6 and 5 dimensions, respectively. 
Assuming that 
 4th (5th) generation vector-like fields only couple to 
 the 1st (2nd) generation chiral fields, 
 the suitable fermion mass hierarchies and flavor mixings 
 are generated by 
 integrating out 
 these vector-like heavy fields. 
The mixing angles between the chiral fields and 
 extra generations have been determined by
 the volume suppression factors. 
The gauge and Higgs fields live in 6 dimensions 
 with the anomaly-free contents, 
 and the orbifolding and boundary conditions make the $SO(10)$ 
 gauge group be broken to 
 $SU(3)_C\times SU(2)_L \times U(1)_Y \times U(1)_X$ and
 realize the TD splitting.

In this paper we will suggest an alternative, much simpler 
 scenario 
 based on the 6D $N=1$ SUSY $SO(10)$ GUT 
 on $T_2/(Z_2 \times Z_2')$.
{}For matter fields, we do not introduce 
 extra vector-like generations
 in extra dimensions as in Ref.~\cite{HKS,HKS2,HS}. 
We will introduce 6D and 5D vector-like matter fields 
 which contains the 1st and 2nd generation matter fields as
 the zero modes, respectively. 
On the other hand, the 3rd generation matter fields are located
 on a 4D brane. 
The Yukawa couplings for the 6D and 5D fields are
 suppressed by volume factors. 
Then, as the result, 
 the suitable fermion mass hierarchies and flavor mixings 
 are generated at low energy.

\section{Fermion mass hierarchies}

We consider the 6D $N=1$ SUSY $SO(10)$ GUT with 
 the vector-like matter contents, whose  
 extra dimensional coordinates are compactified 
 on a $T^2/(Z_2 \times Z_2')$ 
 orbifold\cite{HS}.
The structure of extra 2D spaces are 
 characterized by reflection $P$ $(Z_2)$, $P'$ $(Z_2')$, 
 and translations $T_i$ ($i = 1, 2$). 
Under the reflection $P$ and $P'$, 
 $(z, \bar{z})$ is transformed into 
 $(-z, \bar{z})$ and $(z, -\bar{z})$, respectively,
where $z \equiv (x_5+ix_6)/2$ and $\bar{z} \equiv (x_5-ix_6)/2$
 with the physical space of 
 $0 \leq x_5, x_6 < \pi R$. 
Under the translation 
 $T_1$ and $T_2$, $(z, \bar{z})$ are transformed into 
 $(z + 2\pi R_z, \bar{z})$ and 
 $(z, \bar{z} + 2\pi R_{\bar{z}})$, respectively,
 where $R_z\equiv (1+i)R/2$ and $R_{\bar{z}}\equiv (1-i)R/2$.  
The physical space can be taken as $0 \leq z < \pi R_z$ and 
 $0 \leq \bar{z} \leq \pi R_{\bar{z}}$. 
Thus, the $T_2/(Z_2 \times Z_2')$ orbifold is just 
 the same as the $S_1/Z_2 \otimes S_1/Z_2'$ orbifold
 of a regular square. 
There are four fixed points at 
 $(0,0)$, $(\pi R_z,0)$, $(0, \pi R_{\bar{z}})$ and 
 $(\pi R_z, \pi R_{\bar{z}})$, 
 and two fixed lines on $z=0$ and $\bar{z}=0$ 
 on the orbifold. 
The bulk fields are decomposed by $P$, $P'$, and $T_i$. 
{}For examples, 
 a 6D bulk scalar field $\Phi(x^\mu, z,\bar{z})$ is 
 decomposed into
\begin{eqnarray}
\Phi_{(\pm \pm )(\pm \pm)}(x^{\mu},z,\bar{z}) &\equiv& 
 \frac{1}{\pi R_c}
   \phi_{(\pm \pm)_{z}(\pm \pm)_{\bar{z}}}(x^\mu)
    \varphi_{(\pm \pm)_{z}}(z)
    \varphi_{(\pm \pm)_{\bar{z}}}(\bar{z}), 
\end{eqnarray}
according to the eigenvalues 
 of $(P,T_1)(P',T_2)\; (=(P,T_1)_{z}\otimes(P',T_2)_{\bar{z}})$
,where $R_c \equiv |R_z| (=|R_{\bar{z}}|)$. 
Notice that only $\Phi_{(+\pm )(+\pm )}$ can have massless 
 zero-modes 
 and survives in the low energy.

We consider 
 the gauge multiplet and two ${\mathbf 10}$ representation 
 Higgs multiplets propagate in the 6D bulk,
 which are denoted as ${\bf H_{10}}$ and ${\bf H'_{10}}$.
We adopted the translations 
 as $T_{51} = \sigma_2\otimes I_5$ and 
 $T_{5'1'} = \sigma_2\otimes diag.(1,1,1,-1,-1)$, 
 which commute with the generators
 of the Georgi-Glashow $SU(5)\times U(1)_X$\cite{GG} and 
 the flipped $SU(5)'\times U(1)'_X$\cite{Fl} groups, 
 respectively\cite{ABC,HNOS}.
Then, translations ($T_i$) make the $SO(10)$ 
 gauge group be broken to 
 $SU(3)_C\times SU(2)_L \times U(1)_Y \times U(1)_X$ and
 realize the TD splitting
 since the doublet (triplet) Higgs fields have (not)
 Kaluza-Klein zero-modes.

The zero mode of the 
 6D bulk matter field, 
 ${\psi_{\bf 16}}_{(+\pm )(+\pm)}$,  
 is classified into four types as 
\begin{eqnarray}
 {\psi_{\bf 16}}_{(++)(++)}&~~~&\mbox{(zero mode)~}=Q, \nonumber\\
 {\psi_{\bf 16}}_{(++)(+-)}&~~~&\mbox{(zero modes)}=\overline{U}
,\overline{E}, \nonumber\\
 {\psi_{\bf 16}}_{(+-)(++)}&~~~&\mbox{(zero modes)}=\overline{D},
\overline{N}, \nonumber  \\
 {\psi_{\bf 16}}_{(+-)(+-)}&~~~&\mbox{(zero mode)~}=L. 
\label{6d}
\end{eqnarray}
Similarly, 
 the zero mode of the 5D bulk field, 
 which is propagating on the fixed line $\bar{z} = 0$,  
 is classified into 
\begin{eqnarray}
 {\psi_{\bf 16}}_{(++)}&~~~&\mbox{(zero mode)~}=Q,\overline{U},
\overline{E}, \nonumber\\
 {\psi_{\bf 16}}_{(+-)}&~~~&\mbox{(zero mode)}=L,\overline{D}
,\overline{N}, 
\label{5d}
\end{eqnarray}
where the 2nd $\pm$ sign represents the $T_1$ parity.
\begin{figure}
\begin{center}
\epsfig{file=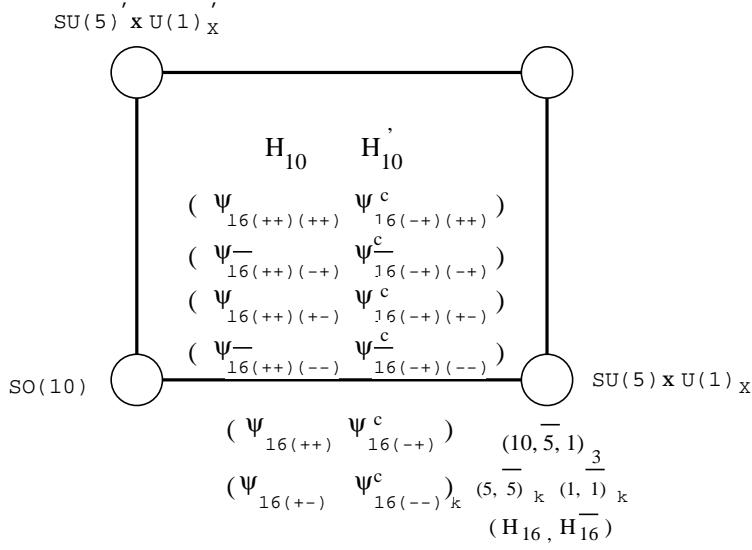,width=10cm}
\caption{Matter configuration on the orbifold. }
\label{fig}
\end{center}
\end{figure}

Now let us discuss matter contents in our model.
In Fig.\ref{fig} we show the the matter configuration
 on the orbifold.
We introduce the 3rd generation matter fields,
 ${\bf 16}_3$ of SO(10), on the 4D brane, $(\pi R_z,0)$.
We also introduce vector-like matter fields,
 ${\bf 5_{(-3)}}_i + {\bf \overline{5}_{(3)}}_i$ and 
 ${\bf 1_{(-5)}}_i + {\bf \overline{1}_{(5)}}_i$
 under $SU(5)\times U(1)_X$, on $(\pi R_z,0)$,
 where $i=1,2$ represents generation index. 
It is possible to put $SU(5)$ multiplets
 on the 4D brane, $(\pi R_z,0)$, since 
 the gauge symmetry 
 is the Gerogi-Glashow $SU(5)\times U(1)_X$
 at this 4D brane. 
On $\bar{z} = 0$, we introduce 5D bulk fields\footnote{The 
 5D bulk fields in $N=1$ SUSY correspond to super-multiplets
 in 4D $N=2$ SUSY.}, 
 $({\psi_{\bf 16}}_{(++)}, {\psi_{\bf 16}}_{(-+)}^c)$, which contains 
 ${\bf 10_{(-1)}}$ of $SU(5)\times U(1)_X$ 
 as the zero mode, and it is regarded as  
 the 2nd generation field. 
We also introduce two 5D bulk fields,
 $({\psi_{\bf {16}}}_{(+-)},{\psi_{\bf {16}}}_{(--)}^c)$ 
 and 
 $({\psi_{\bf {16}}}'_{(+-)},{\psi_{\bf {16}}}'^c_{(--)})$, 
 which contain zero mode components 
 ${\bf \overline{5}_{(-3)}}$ and ${\bf 1_{(-5)}}$ 
 of $SU(5)\times U(1)_X$, and they are 
 denoted as ${\bf \overline{5}}_{\psi_{{\bf {16}}_{i}}}$ 
 and ${\bf 1}_{\psi_{{\bf \overline{16}}_{i}}}$, respectively. 
In 6D bulk, we introduce 6D bulk vector-like fields, 
 $({\psi_{\bf 16}}_{(++)(++)}, {\psi_{\bf 16}}_{(-+)(++)}^c)+
  ({\psi_{\bf \overline{16}}}_{(++)(-+)}, 
   {\psi_{\bf \overline{16}}}_{(-+)(-+)}^c)$ and 
 $({\psi_{\bf 16}}_{(++)(+-)}, {\psi_{\bf 16}}_{(-+)(+-)}^c)+
  ({\psi_{\bf \overline{16}}}_{(++)(--)}, 
   {\psi_{\bf \overline{16}}}_{(-+)(--)}^c)$, 
 which contain the 1st generation ${\bf 10_{(-1)}}$ of 
 $SU(5)\times U(1)_X$ 
 as the zero mode.
Notice that this matter content induces the gauge anomaly 
 neither on the 4D brane nor in 5D and 6D bulks. 
We also introduce the 4D brane-localized 
 Higgs fields, ${H_{\bf 16}}$ and ${H_{\overline{\bf 16}}}$, 
 at $(\pi R_z,0)$, which are assumed to take 
 vacuum expectation values (VEVs) of $O(10^{16})$ GeV 
 in the directions of $B-L$. 
We impose the Peccei-Quinn symmetry and its charge on the multiplets:
The matter fields, ${\bf 5_{(-3)}}_i$, 
 ${\bf \overline{1}_{(5)}}_i$, 
 ${\bf \overline{5}}_{\psi_{{\bf {16}}_{i}}}$, and 
 ${\bf 1}_{\psi_{{\bf \overline{16}}_{i}}}$ 
 have zero,  
 other matter multiplets have $1$, $\mathbf{10}$
 representation Higgs multiplets have $-2$, 
 and $\mathbf{16}$ and $\overline{\mathbf{16}}$ representation 
 Higgs multiplets have $-1$.

The superpotential of  the brane-localized matter fields 
is given by\footnote{Notice that the mass terms 
of vector-like matter fields 
are forbidden by the PQ symmetry.}
\begin{eqnarray}
\label{Yukawa1}
 W_Y = \frac{1}{M_*}\left(
 y^u_{33} \, {\bf 10}_3\, {\bf 10}_3\, {\bf 5_{H_{10}}}
+
\sum_{i=1}^{3}y^u_{3i}\, {\bf \overline{5}}_3\, {\bf 1}_i\,{\bf 5_{H_{10}}}
+
\sum_{i=1}^{3} y^d_{3i}\,  {\bf 10}_3 \,
{\bf \overline{5}}_i\, {\bf \overline{5}_{H'_{10}}}
\right)\delta(z-\pi R_z)\delta(\bar{z}),
\end{eqnarray}
in which $M_*$ is an ultraviolet cut-off scale of $O(10^{18})$ GeV. 
In addition to the superpotential in Eq.(\ref{Yukawa1}),
 there are the following interactions between the brane-localized fields
 and the bulk fields on the 4D brane, $(\pi R_z,0)$, 
\begin{eqnarray}
\label{W6}
W_Y' &=& 
\left\{
\frac{y^u_{11}}{M_*^{3}}\,
{\bf 10}_{\psi_{{\bf 16}_{1}}}\,{\bf 10}_{\psi_{{\bf 16}_{1}}}
 \,{\bf 5_{H_{10}}}
+
\frac{y^u_{22}}{M_*^{2}}\,
{\bf 10}_{\psi_{{\bf 16}_{2}}}\,{\bf 10}_{\psi_{{\bf 16}_{2}}}
 \,{\bf 5_{H_{10}}}
\right.
\nonumber\\
&+&\left.
\frac{y^u_{12}}{M_*^{5/2}}\,
{\bf 10}_{\psi_{{\bf 16}_{1}}}\,{\bf 10}_{\psi_{{\bf 16}_{2}}}
 \,{\bf 5_{H_{10}}}
+
\frac{y^u_{13}}{M_*^{2}}\,
{\bf 10}_{\psi_{{\bf 16}_{1}}}\, {\bf 10}_3 \,{\bf 5_{H_{10}}}
+
\frac{y^u_{23}}{M_*^{3/2}}\,
{\bf 10}_{\psi_{{\bf 16}_{2}}}\, {\bf 10}_3 \,{\bf 5_{H_{10}}}
\right.
\nonumber\\
&+&\left.
\sum_{i=1}^{3}\frac{y^d_{1i}}{M_*^{2}}\,
{\bf 10}_{\psi_{{\bf 16}_{1}}}\, {\bf \overline{5}}_i \,
{\bf \overline{5}_{H'_{10}}}
+
\sum_{i=1}^{3}\frac{y^d_{2i}}{M_*^{3/2}}\,
{\bf 10}_{\psi_{{\bf 16}_{2}}}\, {\bf \overline{5}}_i \,
{\bf \overline{5}_{H'_{10}}}
\right\}
\delta(z-\pi R_z)\delta(\bar{z}).
\end{eqnarray}
%
We identify the zero modes of 
 ${\bf 10}_{\psi_{{\bf 16}_{1}}}$
 and ${\bf 10}_{\psi_{{\bf 16}_{2}}}$ as the 1st and 2nd ${\bf 10}$ 
 massless matter fields, ${\bf 10}_1$ and ${\bf 10}_2$, respectively.
Below the compactification scale, 
 the interactions in Eq.(\ref{W6}) induce the following
 Yukawa couplings for the zero modes.
\begin{eqnarray}
\label{W_4}
W_Y' &=& 
\frac{1}{M_*}\left\{
y^u_{11}\,\epsilon_1^4 \,
{\bf 10}_{1}\,{\bf 10}_{1}
 \,{\bf 5_{H_{10}}}
+
y^u_{22}\,\epsilon_2^2\,
{\bf 10}_{2}\,{\bf 10}_{2}
 \,{\bf 5_{H_{10}}}
\right.
\nonumber\\
&+&\left.
y^u_{12}\,\epsilon_1\epsilon_2\,
{\bf 10}_{1}\,{\bf 10}_{2}
 \,{\bf 5_{H_{10}}}
+
y^u_{13}\,\epsilon_1\,
{\bf 10}_{1}\, {\bf 10}_3 \,{\bf 5_{H_{10}}}
+
y^u_{23}\,\epsilon_2\,
{\bf 10}_{2}\, {\bf 10}_3 \,{\bf 5_{H_{10}}}
\right.
\nonumber\\
&+&\left.
\sum_{i=1}^{3}y^d_{1i}\,\epsilon_1\,
{\bf 10}_{1}\, {\bf \overline{5}}_i \,
{\bf \overline{5}_{H'_{10}}}
+
\sum_{i=1}^{3}y^d_{2i}\,\epsilon_2\,
{\bf 10}_{2}\, {\bf \overline{5}}_i \,
{\bf \overline{5}_{H'_{10}}}
\right\}.
\end{eqnarray}
The 1st and 2nd generation ${\bf 10}$ multiplets of $SU(5)$ give 
the volume suppression factor $\epsilon_1$ and $\epsilon_2$,
respectively, which are given by 
\begin{eqnarray} 
 \epsilon_1 = \epsilon_2 \equiv 1/ \sqrt{{\pi R_c M_*}}. 
\label{eps} 
\end{eqnarray}
These volume suppression factors 
 play crucial roles for generating 
 the fermion mass matrices 
 in the low energy. 
Now we set $1/R_c = O(10^{16})$ GeV, which means 
 $\epsilon_i \simeq\lambda^2 \sim 0.04$,  
 where $\lambda$ is the Cabibbo angle, $\lambda \sim 0.2$.
Although the bulk Higgs fields ${\bf 5_{H_{10}}}$ and 
 ${\bf \overline{5}_{H'_{10}}}$ also induce
 the volume suppression factor, $\epsilon_1$, 
 this effect is not available in the low energy by 
 assuming  
 the original Yukawa couplings $y_u$s, $y_d$s 
 being of $O(\epsilon_1^{-1})$ in 
 Eqs.(4) and (6)\footnote{When we input 
 Higgs fields, ${\bf 5_{H_{(2)}}}$ and 
 ${\bf \overline{5}_{H'_{(-2)}}}$, on 
 the 4D brane, $(\pi R_z, 0)$, 
 there are no volume suppressions for the 
 Higgs fields,
where the original Yukawa couplings $y_u$s, $y_d$s 
 are of $O(1)$ in 
 Eqs.(4) and (6).
In this case, however, TD splitting
 should be solved by other mechanisms.}.

{}For other unwanted matter multiplets, 
 ${\bf 5_{(-3)}}_i$, 
 ${\bf \overline{1}_{(5)}}_i$, 
 ${\bf \overline{5}}_{\psi_{{\bf {16}}_{i}}}$, and 
 ${\bf 1}_{\psi_{{\bf \overline{16}}_{i}}}$,  
 the zero modes become heavy through 
 the mass terms 
\begin{eqnarray} 
\label{W_v}
W_v = \sum_{i=1}^{2}\frac{1}{M_*^{3/2}}\left(
M_{5}^i\, {\bf 5}_i\, {\bf \overline{5}}_{\psi_{{\bf {16}}_{i}}}
+
M_{1}^i\, {\bf \overline{1}}_i \,{\bf 1}_{\psi_{{\bf {16}}_{i}}}
\right)
\delta(z-\pi R_z)\delta(\bar{z}) ,
\end{eqnarray}
on $(\pi R_z, 0)$,
where we set the diagonal bases of 
 $M_5^i$ and $M_1^i$, for simplicity. 
It suggests that these unwanted 
 matter fields decouple in the low energy.
Therefore the particle contents at the low energy become the
 same as those of the minimal SUSY standard model (MSSM)
 with right-handed neutrinos.

Now let us show the mass matrices of the 
 quarks and leptons. 
{}From Eqs.(\ref{Yukawa1}) and (\ref{W_4}),
 the mass matrices of the ordinal matter fields are given by
\begin{equation}
 m_u^l \simeq \left(
\begin{array}{ccc}
 \lambda^8 & \lambda^6 &  \lambda^4  \\
 \lambda^6 & \lambda^4 &  \lambda^2  \\
 \lambda^4 & \lambda^2 &  1
\end{array}
\right)  v, \;\;
 m_d^l \simeq \left(
\begin{array}{ccc}
\lambda^4 & \lambda^4 &  \lambda^4  \\
\lambda^2 & \lambda^2 &  \lambda^2  \\
  1 & 1 & 1 
\end{array}
\right) \overline{v}, \;\;
 m_e^l \simeq \left(
\begin{array}{ccc}
 \lambda^4 & \lambda^2 & 1  \\
 \lambda^4 & \lambda^2 & 1  \\
 \lambda^4 & \lambda^2 & 1  
\end{array}
\right)  \overline{v}, \;\;
\label{mass}
\end{equation}
for the up quark sector, the down quark sector, 
 and the charged lepton sector, respectively \cite{HKS}.
 $\overline{v}$ and    
 $v$ are the vacuum expectation values of the weak Higgs doublets.  
We write the mass matrices in the basis
 that the left-handed fermions are to the left 
 and the right-handed fermions are to the right. 
We notice that all elements in the mass matrices 
 have $O(1)$ 
 coefficients.  
The fermion mass 
 hierarchies are given by
\begin{eqnarray}
\qquad\quad 
m_t : m_c : m_u  &\simeq & 1 : \lambda^4 : \lambda^8\;, \\  
m_b : m_s : m_d  &\simeq& m_{\tau} : m_{\mu} : m_e 
\simeq 1 : \lambda^2 : \lambda^4\;,
\end{eqnarray}
with the large $\tan \beta$. 
The mass matrix of three light neutrinos $m_\nu^{(l)}$ through
 the see-saw mechanism\cite{seesaw} is given by 
\begin{equation}
\label{26}
 m_\nu^{(l)} \simeq
 {m_\nu^D m_\nu^D{}^T \over M_R} \simeq 
\left(
\begin{array}{ccc}
 1 & 1 & 1 \\
 1 & 1 & 1 \\
 1 & 1 & 1 
\end{array}
\right)  {v^2 \over M_R}.
\end{equation}
$M_R$ is about 
$10^{14}$ GeV induced from the interaction 
\begin{equation}
\label{Yukawa2}
W_{M_N} = \frac{y^{N}_{ij}}{M_*} \overline{1}_{H_{\overline{16}}}
\overline{1}_{H_{\overline{16}}} {\bf 1}_i {\bf 1}_j 
\delta(z-\pi R_z)\delta(\bar{z})
\end{equation}
on $(\pi R_z,0)$. 
We can obtain the suitable mass scale ($O(10^{-1})$ eV) for
 the atmospheric neutrino oscillation experiments, 
 by taking account of the $SO(10)$ relation, 
 $y_u \simeq y_{\nu}$.

As for the flavor mixings, 
 the CKM\cite{CKM} and the MNS\cite{MNS} matrices are given by 
\begin{equation}
\label{mixing1}
 V_{CKM} \simeq \left(
\begin{array}{ccc}
 1 & \lambda^2 & \lambda^4 \\
 \lambda^2 & 1 & \lambda^2 \\
 \lambda^4 & \lambda^2 & 1 
\end{array}
\right)  , \;\;
 V_{MNS} \simeq \left(
\begin{array}{ccc}
 1 & 1 & 1 \\
 1 & 1 & 1 \\
 1 & 1 & 1 
\end{array}
\right) . 
\end{equation}
which realize the suitable flavor mixings 
 roughly in order of magnitudes. 
They give us a natural explanation 
 why the flavor mixing in the quark sector 
 is small while the flavor mixing in the 
 lepton sector is large\cite{BB}-\cite{anarchy}.
{}For the suitable values of 
 $V_{us}$, $U_{e3}$, and 
 down quark and electron masses, 
 we need suitable choice of $O(1)$ coefficients 
 in mass matrices as in Ref.~\cite{BB}. 
Or, if $O(1)$ coefficients are not determined by 
 a specific reason (symmetry) in the fundamental theory, 
 it is meaningful to 
 see the most probable hierarchies and mixing angles 
 by considering random $O(1)$ coefficients\cite{anarchy}.
Anyway, if the fermion mass hierarchies and 
 flavor mixing angles should be determined from the fundamental 
 theory 
 in {\it order} (power of $\lambda$) not by tunings of 
 $O(1)$ coefficients, 
 we should modify our mechanism. 
The modification in our scenario 
 can be achieved in a simple way. 
We introduce a $({\psi_{\bf {16}}}_{(+-)},{\psi_{\bf {16}}}_{(--)}^c)$ 
 matter hyper-multiplet with PQ-charge 1 on $\bar{z}=0$, 
 instead of the brane-localized ${\bf \overline{5}}_1$. 
By identifying 
 the zero mode of this matter field 
 as the 1st generation ${\bf 5}$ of $SU(5)$, 
 the 1st generation ${\bf 5}$ induces a volume suppression
 factor, $\lambda^2$, similar to the case of ${\bf 10}_{1,2}$
 representations. 
By this modification, we can obtain 
 the modified mass matrices of quarks and leptons, 
 which can induce the suitable values of 
 $V_{us}$, $U_{e3}$, and 
 down quark and electron masses
 (this type of mass matrices is the 
 modification I in Ref.~\cite{HKS2}). 
Furthermore, when we introduce 
 $({\psi_{\bf 16}}_{(+-)(++)}, {\psi_{\bf 16}}_{(--)(++)}^c)+
  ({\psi_{\bf \overline{16}}}_{(+-)(-+)}, 
   {\psi_{\bf \overline{16}}}_{(--)(-+)}^c)$ and 
 $({\psi_{\bf 16}}_{(+-)(+-)}, {\psi_{\bf 16}}_{(--)(+-)}^c)+
  ({\psi_{\bf \overline{16}}}_{(+-)(--)}, 
   {\psi_{\bf \overline{16}}}_{(--)(--)}^c)$,  
 in 6D bulk, and two 
 $({\psi_{\bf {16}}}_{(+-)},{\psi_{\bf {16}}}_{(--)}^c)$s 
 on $\bar{z}=0$ (these bulk fields contain ${\bf 5}$ matter fields 
 of $SU(5)$ as the zero modes), and 
 identify the zero modes ${\bf 5}$ fields as 
 1st and 2nd, 3rd generations, 
 respectively, 
 we can obtain mass matrices of the small $\tan \beta$ case 
 in the similar manner (modification II in Ref.~\cite{HKS2}).

\section{Summary and Discussion}

In this 
 paper, we have shown the models based on 
 the 6D $N = 1$ SUSY $SO(10)$ GUT 
 where the 5th and 6th dimensional coordinates are 
 compactified on a $T^2/(Z_2 \times Z_2')$ orbifold. 
The gauge and Higgs fields live in 6 dimensions.
The TD splitting and the gauge symmetry reduction are realized by
 the orbifolding. 
The matter fields are contained with the anomaly-free
 contents. 
We introduce the 6D and 5D matter fields, 
 which contains the 1st and 2nd generation fields 
 as the zero modes, respectively. 
The Yukawa couplings for the bulk fields are suppressed by
 the volume factors. 
The fermion mass hierarchy 
 has been realized, since the 1st (2nd) generation fields
 propagate in 6D (5D). 
We have shown that the suitable fermion mass hierarchies 
 and flavor mixings are generated by the volume suppression 
 factors originated from extra dimensions.

Let us briefly discuss the SUSY breaking mechanisms. 
In the gaugino mediation scenario \cite{KKS}, 
the SUSY breaking field, $S=\theta^2F$,
are located at a different fixed point from
our brane, $(\pi R_z,0)$. Then the gauginos obtain
the SUSY breaking masses from the interaction with
$S=\theta^2F$. 
The SUSY breaking for
 brane-localize matter fields 
 are generated through the quantum effects of 
 the gaugino masses.
While the matter fields in extra dimensions
 can directly couple to the SUSY breaking fields,
 where non-universal contributions to
 SUSY breaking masses can be generated in general. 
Since these non-universal SUSY breaking masses can give rise to
 too large flavor changing neutral currents (FCNCs),  
 the location of the SUSY breaking brane 
 should be determined in order to avoid the large FCNC 
 phenomenological problems in the gaugino mediation 
 scenario\cite{HS}. 
On the other hand, 
 in the gauge mediation scenario, 
 we can put the messenger sector on the 4D brane,
 $(\pi R_z, 0)$, as in Ref.~\cite{HS}. 
The SUSY breaking masses for the light matter fields 
 are highly degenerated where the FCNCs
 are naturally suppressed as in the ordinal 
 4D gauge mediation models. 
For other 
 SUSY breaking 
 scenarios,  
 the non-universal corrections to the 
 soft SUSY masses can be arisen in
 the gravity mediation scenario in general. 
It is because the SUSY breaking effects 
 are mediated by ``Yukawa'' interactions not 
 by gauge interactions. 
The ``Yukawa'' interactions 
 among the bulk fields always receive 
 the volume suppressions, which 
 violate the degeneracy of the soft SUSY breaking masses. 
The Scherk-Schwarz SUSY breaking \cite{SS,SS2} 
 might also give the non-negligible effects 
 of breaking degeneracy, since 
 the first and the second generation fields
 are mainly composed by the bulk fields in 
 our models.


\section*{Acknowledgment}
We would like to thank A. Sugamoto for helpful discussions. 
This work is supported in
 part by the Grant-in-Aid for Science
 Research, Ministry of Education, Culture, Sports, Science and 
Technology, of Japan (No. 14039207, No. 14046208, No. 14740164).


\end{document}